 \documentclass[final,5p,times,twocolumn]{elsarticle}

\usepackage{amssymb}
\usepackage{graphicx}
\usepackage{dcolumn}

\journal{Physica C}

\begin{document}

\begin{frontmatter}



\title{Electron-Phonon Properties of Pnictide Superconductors}

\author{L. Boeri}
\address{Max-Planck-Institut f\"{u}r Festk\"{o}rperforschung,
Heisenbergstra$\rm \beta$e 1, D-70569 Stuttgart, Germany}
\author{O.V. Dolgov}
\address{Max-Planck-Institut f\"{u}r Festk\"{o}rperforschung,
Heisenbergstra$\rm \beta$e 1, D-70569 Stuttgart, Germany}

\author{A.A. Golubov}
\address{Faculty of Science and Technology and MESA+ Institute for Nanotechnology,
University of Twente, 7500 AE Enschede, The Netherlands}

\date{\today}


\begin{abstract}
In this paper we discuss the normal and superconducting state properties of two pnictide superconductors,
LaOFeAs and LaONiAs, using Migdal-Eliashberg theory and density functional perturbation theory.
For pure LaOFeAs, the calculated electron-phonon coupling constant $\lambda=0.21$
and logarithmic-averaged frequency
$\omega_{ln}=206 K$, give a maximum $T_c$ of 0.8 K, using the standard
Migdal-Eliashberg theory. Inclusion of multiband effects increases the Tc only marginally.
To reproduce the experimental $T_c$, a 5-6 times larger coupling constant
would be needed.
Our results indicate that standard electron-phonon coupling is not sufficient to explain superconductivity
in the whole family of Fe-As based superconductors.
At the same time, the electron-phonon coupling in Ni-As based compounds is much stronger
and its normal and superconducting state properties can be well described by standard Migdal-Eliashberg theory.
\end{abstract}


\end{frontmatter}


\section*{Introduction}
\label{intro}
Strong-coupling electron-phonon (EP) theory, also known as Migdal-Eliashberg
(ME) theory, was developed in the 60's and 70's to describe the physical
properties of superconducting elemental metals and alloys, which could
not be described by the weak-coupling BCS approach.

The electronic and phononic systems are described by a set of coupled
diagrammatic equations (ME equations), which give a complete
description of the normal and superconducting state, including the
superconducting critical temperature.
ME theory has been generalized to include multi-band and anisotropic coupling,
magnetic and non-magnetic impurities, etc.; a review can be found in \cite{review}.

The biggest difficulty in the 60's and 70's was extracting the $EP$ coupling
spectral function from the available experimetal data, which involved some degree of  approximation.
In the last few years, it has become possible to calculate it completely
{\em ab-initio}~\cite{DFT}.
The combination of {\em ab-initio} calculations and ME theory has permitted to calculate the superconducting and normal state properties of many new and old materials with considerable accuracy~\cite{SC:savrasov:band}:
the biggest success is probably
represented by the two-gap superconductor MgB$_2$, with the record $T_c$ of
40 K~\cite{mgb2,mgb2a,mgb2b}.
These methods, however, fail dramatically in more exotic superconductors,
such as the high-$T_c$ cuprates, where the key approximations (weak electronic
correlations, well separated phonon and electronic energy scales) break down
~\cite{HTSC:DFT:savrasov,eph:theory:gunnarsson}.

In this paper, we analyze the possibility of applying ME theory to two
newly-discovered pnictide superconductors, LaOFeAs and LaONiAs.
The motivation  of the application of the ME approach to Fe pnictides is
connected  with the rather large mass  renormalisation ($\lambda \approx 1-1.5$ in
ARPES and de Haas-van  Alphen effects, and slightly smaller
$\lambda_{tr} \approx 0.5$ in transport properties) as well as observed large isotope
shift $\alpha_{Fe}\sim 0.4$ \cite{liu}.

The results for the Fe compound were already presented in our previous
publication~\cite{LFAO:DFT:boeri}, and we review them here, considering also the effect of
multi-band coupling.
We also decided to include new results for the Ni compound, LaONiAs,
which is the only other member of the LaO$M$As family ($M$=Mn,Fe,Co,Ni,Cu,Zn)
~\cite{LFAO:syn1:zimmer}
showing superconductivity, albeit with a much lower $T_c \approx$ 2.4-3.8 $K$
~\cite{LNAO:tc:watanabe,LNAO:exp:li,LNAO:exp:fang}.

Experimental results~\cite{LNAO:tc:watanabe,LNAO:exp:li,LNAO:exp:fang,BN2A2:exp:ronning,LNPO:exp:watanabe,BN2P2:exp:terashima} 
and calculations on other Ni pnictides~\cite{LNPO:DFT:subedi,BN2A2:DFT:subedi} strongly
suggested that this may be a standard $EP$ superconductor; in this work,
we compare the available experimental data for LaONiAs ($T_c$, specific heat,
dHvA) with ME calculations, and we show that there is indeed a very good agreement.

The comparison of these results with those for LaOFeAs is very instructive.
LaOFeAs has in fact a much smaller coupling constant $\lambda$,
which is a factor 5 too low to reproduce the experimental $T_c=26 K$
~\cite{LFAO:tc:kamihara}, even
considering multiband effects. Similarly to the superconducting cuprates,
LaOFeAs and Fe pnictides in general are much more ``exotic'' materials,
where many-body effects may play an important role.

The exotic features of Fe pnictides (itinerant magnetism, structural transitions, unusual gap symmetry, \textit{etc}.) 
are reviewed in detail in other contributions
to this issue, and we will not treat them here, although in the last part of this paper we will discuss how they can affect our results.

This paper is organized as follows: In section ~\ref{electrons} we present
the band structure and Fermi surface of the two compounds; in section
~\ref{phonons} we show the phonon dispersions and electron-phonon coupling
calculated in Linear Response theory; in section ~\ref{ME}
we present the Migdal-Eliashberg results; in the last section
we discuss our results in light of our experimental and theoretical
works. The technical details of the density functional theory (DFT) calculations are given in Appendix ~\ref{details}.

\section{Electronic Structure}
\label{electrons}
\begin{figure}
\begin{center}
\includegraphics*[width=5cm]{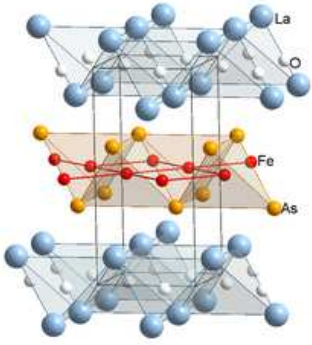}
\end{center}
\caption{\label{fig:fig1}(color online) Crystal structure of LaOFe(Ni)As, from Ref.~\cite{LFAO:DFT:boeri}.}
\end{figure}
LaOFeAs and LaONiAs crystallize
 in the ZrCuSiAs structure (space group 129);
the primitive cell is tetragonal,
La and As atoms occupy $2c$ Wyckoff positions,
O and $M$ atoms ($M$=Fe,Ni) occupy $2a$ and $2b$ Wyckoff positions.

The structure, depicted in Fig.~\ref{fig:fig1}, consists of alternating
$M$-As and La-O layers.
$M$ and O atoms sit at the
center of slightly distorted As and La tetrahedra;
the As tetrahedra are squeezed in the $z$ direction, so that
there are  two $M-As-M$ angles ($\theta_1$, $\theta_2$),
which are either larger or smaller than the regular tetrahedron value
($\theta_0=109.47 \deg$).
$M$ atoms form a square lattice; the $M-M$ in-plane distance
is $\sim 20 \%$ larger than the $M$-As one.
The relevant parameters of the structure for the two compounds are given in
Table~\ref{table:struct}.
\begin{center}
\begin{table*}[h!tbp]
{\small
\hfill{}
\begin{tabular}{|l|c|c|c|c|c|c|c|c|c|c}
\hline
              & $a$         & $c$    & $z_{As}$&   $z_{La}$ & d$_{As-M}$ &
d$_{M-M}$ & $\theta_1$ & $\theta_2$ \\
\hline
LaOFeAs (exp)& 4.035 & 8.741 & 0.6512 & 0.1415 & 2.41 & 2.85 & 107.5 &113.5 \\
LaOFeAs (th)& 3.996 & 8.636 & 0.6415 & 0.1413 & 2.34 & 2.83 & 105.81 &117.1 \\
LaONiAs (exp)& 4.123 & 8.1885 & 0.6368 & 0.1470 & 2.35 & 2.92 & 103.19 & 122.95 \\
LaONiAs (th)& 4.102 & 8.2886 & 0.6398 & 0.1423 & 2.36 & 2.90 & 103.99 &121.11 \\
\hline
\end{tabular}}
\hfill{}
\caption{Structural data of LaOFeAs and LaONiAs from experiment (Refs. ~\cite{LFAO:tc:kamihara} and~\cite{LNAO:tc:watanabe}), 
and DFT (Ref.~\cite{LFAO:DFT:boeri} and this work). Distances are in $\AA$, angles in
  degrees; for a perfect tetrahedron, $\theta_1$=$\theta_2$=109.47 $\deg$.}
\label{table:struct}
\end{table*}
\end{center}
In this table we report both the experimental data,
from Ref.~\cite{LNAO:tc:watanabe,LFAO:tc:kamihara}, and the data that we obtained from a full
DFT structural relaxation (Ref.~\cite{LFAO:DFT:boeri} and present work),
which we will use in the following calculations.
As it was noticed by several authors in literature, in the Fe compound
non spin-polarized
DFT calculations
tend to strongly overestimate the $As$
tetrahedron deformation with respect to the experiment;
the agreement is
improved if spin polarization is allowed~\cite{LFAO:DFT:yin},
which is normally interpreted as
a sign of spin fluctuations~\cite{LFAO:DFT:mazin2,LFAO:DFT:mazin3}.
On the other hand, in the Ni compound,
the tetrahedral angles given by non-spin polarized DFT
calculations are very close to those
found experimentally (see Table~\ref{table:struct}).

Our band structures of LaOFeAs
and LaONiAs are in very good agreement with literature results
~\cite{LFAO:DFT:boeri,LFAO:DFT:yin,LFAO:DFT:mazin,LFAO:DFT:singh,LNAO:DFT:xu}.
The most important difference between the two compounds is a $\sim 1$ eV
shift of the Fermi level of the Ni ($d^8$) with respect to the Fe ($d^6$)
compound, due to the different electron count.
%
%
Measuring energies from the Fermi level LaOFeAs compound,
O $p$ and As $p$ states form a group of 12 bands extending from
$\sim -6$ to $-2$ eV.
La-$f$ states are found at higher energies, at $\sim 2$  eV.
The dominant contribution to the states in
 an energy window extending $\pm 2$ eV
around the Fermi level comes from the ten M$-d$ states, which
hybridize with the As $p$ states.

\begin{figure}
\includegraphics*[width=7.8cm]{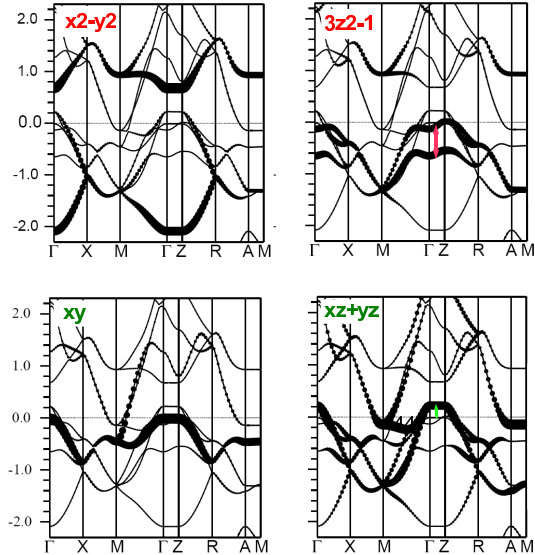}
\caption{\label{fig:fatFe}
Blow-up of the band structure of {\bf LaOFeAs} around the Fermi level,
decorateed with partial $e$ ({\em red}) and $t_2$
({\em green}) Fe characters. From Ref.~\cite{LFAO:DFT:boeri}.
}
\end{figure}

\begin{figure}
\includegraphics*[width=7.8cm]{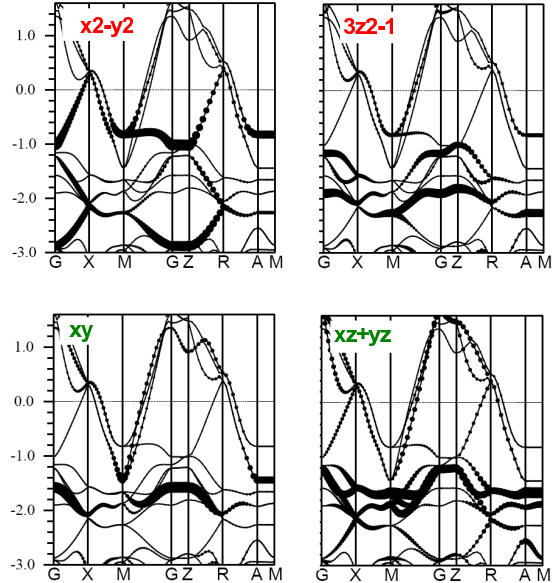}
\caption{\label{fig:fatNi}
Blow-up of the band structure of {\bf LaONiAs} around the Fermi level,
decorateed with partial $e$ ({\em red}) and $t_{2}$
({\em green}) Ni characters.
}
\end{figure}

A blow-up of the band structure in this energy region,
decorated with partial $M$ $d$ character, is shown in
Figs.~\ref{fig:fatFe}-\ref{fig:fatNi}.
The $x,y$ axes are oriented along the $M-M$ bonds.
Due to the strong hybridization with As $p$ states, the
 $d$ bands do not split simply into a lower $e$ ($d_{x^2-y^2}$
and $d_{3z^2-1}$) and higher $t_2$ manifold, as predicted by crystal
field theory.
The
$d_{x^2-y^2}$ orbitals, which lie along the $M-M$ in-plane
bonds, due to hybridization are split into two subsets of flat bands of
located at $-2$ and $+1$ eV.
The $d_{3z^2-1}$ bands have the most three-dimensional character and
sit just below $E_F$.
The $t_{2}$ bands, derived from
$d_{xy}$,$d_{xz}$,
and $d_{yz}$ states,
form a complicated structure centered at $\sim -0.5$ eV, and
give the largest contribution to the Density of States (DOS) at the Fermi level.

The Fermi level of LaOFeAs cuts the band structure in a region where the
DOS is high ( 2.1 states/eV spin) and rapidly decreasing;
a pseudogap opens in the electronic spectrum around $0.2$ eV.
The resulting Fermi surface comprises two  cylindrical hole
pockets centered at the $\Gamma$ point, and a doubly-degenerate electron
pocket centered at the $M$ point; these sheets
have a dominant $d_{xz},d_{yz},d_{yz}$ character.

The quasi-nesting between the hole and electron pockets
leads to a peak in the magnetic susceptibility,
and hence to an instability of the non-magnetic solution
with respect to a striped antiferromagnetically (AFM) ordered phase.
A third hole pocket centered around the $\Gamma$ point is also present;
its character ($d_{3z^2-1}$
or $d_{xy}$) depends on the details of the calculations, and
in particular on the deformation of the As tetrahedra~\cite{LFAO:DFT:mazin2}.
The plasma frequencies are strongly anisotropic
($\omega^{pl}_{xy}=2.30, \omega^{pl}_{z}=0.32$ eV).

A similar blow-up of the band structure for the Ni compound
is shown in Fig.~\ref{fig:fatNi}.
Due to the different electron count, the $xz,yz$ hole pockets are
completely full, and the Fermi surface contains two electron and one hole
sheets, with a marked 2D character.
Besides the elliptical $xz,yz$ pocket,
the second large electron sheet centered at the $M$ point has a
dominant $x^2-y^2$ character; the same bands also forms small hole
pockets around the $X$ point of the Brillouin zone.
The corresponding bands account for the directional
in-plane bonds of the $M$ planes.

The DOS is lower than in the Fe compound,
$N(0)=1.66 st./ eV$ spin, flat, and roughly
particle-hole symmetric in an energy interval corresponding
to 10 $\%$ hole and electron doping.
The Fermi velocities are on average
higher than in LaOFeAs, and the resulting plasma frequency are larger and
strongly anisotropic ($\omega^{pl}_{xy} =4.49$ eV, $\omega^{pl}_{z}=0.45$ eV)).

The $\Gamma$-centered $xz,yz$ hole pockets are
completely full, and this suppresses the tendency to AFM order found in
the Fe compound. In fact, we do not find any AFM solution,
neither in the LSDA nor in the GGA, in agreement with previous
calculations~\cite{LNAO:DFT:xu}.
%
%
\section{Electron-Phonon Properties}
\label{phonons}
\begin{center}
\begin{table*}[h!tbp]
{\small
\hfill{}
\begin{tabular}{|l|c|c|c|c|c|c|c|c|}
\hline
              & $N(0)$  $eV^{-1} f.u.^{-1}$      & $\omega^{pl}_{xy}$ ($eV$)   &   $\omega^{pl}_{z}$ ($eV$) & $\omega_{ln}$ ($K$) & $\lambda$ & $\gamma_0$ ($mJmol^{-1}K^{-2}$) &
  $T_c^{th}$ ($K$) & $T_c^{exp}$ ($K$) \\
\hline
LaOFeAs        & 2.1   & 2.30   & 0.23   & 205    & 0.21 & 4.95      & 0.0 (0.0)     & 26  ~\cite{LFAO:tc:kamihara} \\
LaONiAs        & 1.64  & 4.49   & 0.45   & 96    & 0.72 & 3.86       & 2.9 (3.8)   & 2.4 ~\cite{LNAO:tc:watanabe}, 3.8 ~\cite{LNAO:exp:li}\\
\hline
\hline
FeSe           & 1.9   & -      & -      & 163    & 0.17 & 4.48     & 0.0     & 18  \\
LaONiP         & 1.41  & -      & -      & 162    & 0.58 & 3.32     & 2.6      &  3 \\
BaNi$_2$As$_2$ & 1.78  & -      & -      & 105    & 0.76 & 4.20     & 3.8      &0.7 \\
\hline
\end{tabular}}
\hfill{}
\caption{Electron-phonon properties of Fe and Ni superconductors calculated from Density functional perturbation theory.
The results in the first two rows are from Ref.~\cite{LFAO:DFT:boeri} and this work.
The $T_c$ values in the first column ($T_c^{th}$) are obtained using $\mu^*=0.12$
and Allen-Dynes formula (Eq.~\ref{eq:tc}); number in parentheses correspond to the full numerical solution of
Migdal-Eliashberg equations, given in Sect.~\ref{ME}.
For comparison, in the last three rows we also report literature data on FeSe,
LaONiP and BaNi$_2$As$_2$ from Refs.~\cite{FS:DFT:subedi,LNPO:DFT:subedi,BN2A2:DFT:subedi}.}
\label{table:eph}
\end{table*}
\end{center}
The main results of the linear response calculations for LaOFeAs and LaONiAs are
shown in Table~\ref{table:eph}.
In both cases we performed the calculations at zero doping in the non-magnetic
(NM) phase.
LaONiAs is non-magnetic and superconducting at zero doping; on the other
hand, the ground state of undoped LaOFeAs, both in DFT~\cite{LFAO:DFT:yin}  and experiment, is a
striped antiferromagnetic order; doping suppresses magnetism and leads to superconductivity.
Our NM calculations are thus meant as a model for doped,
superconducting LaOFeAs, considering that
the effect of doping in the virtual crystal approximation is roughly a
rigid-band shift of the Fermi energy ($E_F$),
which does not change the topology of the Fermi surface, but only the
value of the DOS at $E_F$.
It is important to point out that
dynamic spin fluctuations, which have been argued to be present also
in the superconducting Fe samples, are  not included in this calculation, but would
require going beyond the DFT level.
We will discuss this issue in more detail in the final section of this work.

\begin{figure}[h!tbp]
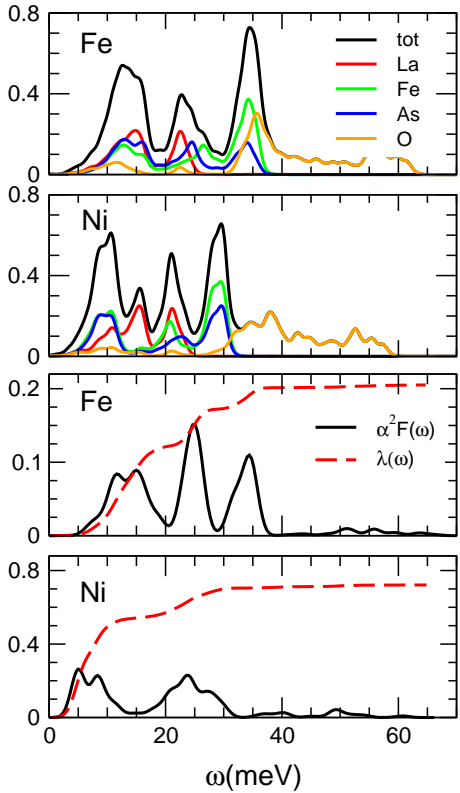

\begin{center}
\includegraphics*[width=6.0cm]{phdos_both.eps}
\includegraphics*[width=6.0cm]{alpha_both.eps}
\end{center}
\caption{\label{fig:figeph}(color online){\em From top to bottom:}
Partial PDOS (Fe,Ni), Eliashberg functions and frequency-dependent
electron-phonon coupling constant $\lambda (\omega)$ (Eq.~\ref{eq:eq1}-
\ref{eq:eq2}) for LaOFeAs and LaONiAs, calculated within Density Functional
Perturbation Theory (DFPT).
}

\end{figure}

In the two top panels of Fig.~\ref{fig:figeph}
we show the atom-projected phonon DOS (PDOS) of LaOFeAs ({\em top})
and LaONiAs ({\em bottom}).
Both spectra extend up to 65 meV, and have a rather similar shape.
The vibrations of O atoms are well
separated in energy from those of other atomic species, lying at $\omega >$ 40 meV.
The vibrations of La, Fe(Ni) and As occupy the same energy
range, and the eigenvectors have a strongly mixed character.
Similarly to the electronic bands, the phonon branches ({\em not shown})
have very little dispersion in the $z$ direction.
Analyzing the evolution of the phonon eigenvectors in the Brillouin Zone (BZ)
reveals that there is no clear
separation between in and out-of-plane vibrations, as it often happens in layered compounds.
The three major peaks in the PDOS at $\omega = 10,20 $ and $30$ meV
do not show a definite in-plane or out-of-plane character, and cannot be easily traced back to
a single vibration pattern.
Their energy is shifted down by $\sim 20$ $\%$ when going from Fe to Ni,
mostly due to EP softening.

The EP coupling of LaONiAs ($\lambda=0.72$) is in fact much larger than in
LaOFeAs ($\lambda=0.21$),
as shown in the two lower panels of Fig.~\ref{fig:figeph}.
Here we plot the two Eliashberg spectral functions $\alpha^2 F(\omega)$,
 together with the
frequency-dependent EP coupling function
$\lambda(\omega)$:
\begin{eqnarray}
\alpha^2 F(\omega)&=&\frac{1}{N\left( 0\right) }\sum_{nm\mathbf{k}}\delta (\varepsilon _{n\mathbf{%
k}})\delta (\varepsilon _{m\mathbf{k+q}}) \times
\nonumber
\\
&\times&
\sum_{\nu \mathbf{q}}|g_{\nu ,\,n
\mathbf{k},\,m\left( \mathbf{k+q}\right) }|^{2}\delta (\omega -\omega _{\nu
\mathbf{q}});
\label{eq:eq1}
\\
\lambda(\omega)&=&2\int_{0}^{\omega } d\Omega \alpha ^{2}F(\Omega )/\Omega,
\label{eq:eq2}
\end{eqnarray}
where  $g_{\nu ,\,n\mathbf{k},m \mathbf{k}'}$
are the bare DFT matrix elements, {\em i.e.} they do not include many-body effects.
A comparison of the Eliashberg function with the PDOS shows that,
apart from the fact that in  both systems
the high-lying O modes
show very little coupling to electrons,
%
there are important differences in the shape and size
of $\alpha^2 F(\omega)$ between the Fe and the Ni compound.
In fact, in LaOFeAs there is an almost perfect proportionality between the
PDOS and the $\alpha^2 F(\omega)$, whereas in LaONiAs the coupling
to the two lowest peaks of the PDOS is strongly enhanced.

A perfect proportionality between the Eliashberg function and the
PDOS (LaOFeAs) implies that there are no patterns of vibration
with a dramatic effect on the electronic states at the Fermi level.
In good EP superconductors, on the other hand, the
coupling to electrons is usually concentrated in a few selected phonon modes.
This is best explained in terms of phonon patterns that awake dormant EP
interaction between strongly directed orbitals.~\cite{A15:DFT:weber}.
This is what happens in LaONiAs, where the electronic states derived
from the Ni $d_{x^2-y^2}$ orbitals sit at the Fermi level, and experience
a strong coupling to the low-energy Ni-As modes.

The different EP coupling of LaONiAs and LaOFeAs derives
from the character of the electronic states at the Fermi energy. Therefore,
it should be a rather general property of the Fe and Ni families of pnictide
superconductors which, apart from minor differences due to
chemistry and structure, share the same band structure.

In fact, as we show in Table \ref{table:eph}, the calculated EP coupling
constants in Ni compounds ($\lambda \approx 0.58-0.76$)~\cite{LNPO:DFT:subedi,BN2A2:DFT:subedi}
 are always 3-4 times larger
than  in Fe-based materials ($\lambda \approx 0.17-0.21$)~\cite{LFAO:DFT:boeri,LFAO:DFT:mazin,FS:DFT:subedi}.
This has important implications on the possible pairing mechanism for
superconductivity in the two classes of materials.
We can get an estimate of $T_c$ due to EP coupling
using Allen-Dynes formula~\cite{Tc:allendynes:theory}:
\begin{equation}
T_c=\frac{\langle\omega_{ln}\rangle}{1.2} \exp \left[
\frac{-1.04(1+\lambda)}{\lambda-(1+0.62\lambda)\mu^*} \right],
\label{eq:tc}
\end{equation}
For $\mu^* = 0.12$, this gives $T_c < 0.01 $ K ($\omega_{ln} = 205$ K) for LaOFeAs
and $T_c = 2.9$ K for LaONiAS ($\omega_{ln} = 96 $ K).

To reproduce the experimental ($T_c=26 K$) of LaOFeAs, a five times larger
$\lambda$ would be needed,
even for $\mu^*=0$.
On the other hand, the DFPT results seem to nicely explain the $T_c$ of
LaONiAs.

In the next section, we will calculate in more detail the normal and
superconducting state properties of LaOFeAs and LaONiAs using the full
ME theory, and compare the results with available experimental data.

\section{Superconducting Properties}
\label{ME}
a) $LaFeAsO_{0.9}F_{0.1}$

In principle, multiband and/or anisotropic coupling could provide the missing
factor 5 in the coupling missing to explain $T_c$ similar to the multiband superconductivity in
MgB$_{2}$ (see, e.g., Refs.\cite{mgb2a,mgb2b});
however this is very unlikely because this would require a very
anisotropic distribution of the EP coupling~\cite{Tc:dolgov:multi}.
Definitely, the iron pnictides are multiband
superconductors with hole and electron bands which are
well separated in momentum space.
To analyze this possibility we split the electron-phonon interaction (EPI)
in Eq.~\ref{eq:eq1}
over electron and hole pockets on the Fermi surface,
obtaining the band-decomposed superconducting Eliashberg functions:
\begin{eqnarray}
\alpha _{ij}^{2}(\omega )F_{ij}(\omega )&=&\frac{1}{N_{i}(0)}\sum_{{\bf k,k}%
^{\prime },\nu }\left| g_{{\bf k,k}^{\prime }}^{ij,\nu }\right| ^{2}
\label{eliash:sc}
\\
&\times&\delta
(\varepsilon _{{\bf k}}^{i})\delta (\varepsilon _{{\bf k^{\prime }}%
}^{j})\delta (\omega -\omega _{{\bf k-k^{\prime }}}^{\nu }),
\nonumber
\end{eqnarray}
where $N_{i}(0)$ is the partial DOS per spin at the Fermi
energy of the $i$'th sheet of the Fermi surface, $g_{{\bf k,k}^{\prime
}}^{ij}$ is the EPI matrix element.
These functions, shown in Fig.~\ref{fig:decomp}, determine the
superconducting properties and thermodynamical properties like electronic
specific heat and de Haas-van Alphen mass renormalizations.
In contrast to MgB$_{2}$, in LaOFeAs the
matrix of coupling constant is practically uniform:
\begin{figure}
[ptb]
\begin{center}
\includegraphics[
height=2.5607in,
width=3.3217in
]%
{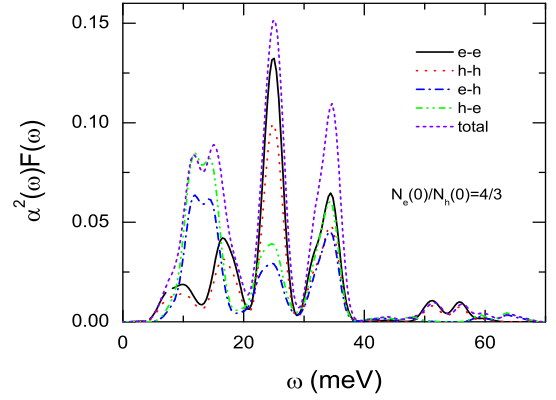}%
\caption{The decomposition of the Eliashberg function on the\textit{
intraband} and \textit{interband} interactions; the ratio between hole and
electron DOS is obtained from a $2^{nd}$ order fit of the band structure aroun $E_F$.}%
\label{fig:decomp}%
\end{center}
\end{figure}
\[
\lambda=\left(
\begin{array}
[c]{cc}%
0.111 & 0.093\\
0.124 & 0.083
\end{array}
\right)  ,
\]
while the characteristic logarithmic frequencies are different ( $\omega_{\ln
}^{intra}=214$ K, and $\omega_{\ln}^{inter}=180$ K). The small
difference in the elements of the matrix $\lambda$ leads to a non-drastic
difference between the maximal eigenvalue and the EPI averaged over bands. As
a result we get slightly larger value of $T_{c}\approx1.5$ K, for $\mu^*=0$
which is much lower than the observed value $T_c=26$ K.
In view of the above result, we do not pursue the ME study of LaOFeAs further.
Other interactions, repulsive in the $s$-wave channel but attractive in the
$d$- or $p$-wave one (\textit{e.g.} spin-fluctuations or the direct Coulomb
interaction), may increase $T_{c}$~\cite{LFAO:DFT:mazin}.

b) $LaNiAsO_{0.9}F_{0.1}$

LaONiAs is superconducting with $T_c \approx 2.4-3.8$ K when hole (Sr) or electron
(F) doped. The phase diagram is roughly symmetric, in agreement
with the flat DOS predicted by DFT in this doping interval.
In this paragraph, we apply ME theory to
LaNiAsO$_{0.9}$F$_{0.1}$, for which we could find
the most complete set of experimental data in literature~\cite{LNAO:exp:li}.

The calculated Eliashberg spectral function $\alpha^{2}(\omega)F(\omega)$
shown in Fig.~\ref{fig:figeph}, with a total $\lambda=0.72$,
yields the experimental $T_{c}=$3.8 K,
with a Coulomb pseudopotential of $\mu^{\ast}=0.12$
( slightly higher then by using the Allen-Dynes expression).
We fix $\mu^*=0.12$ in the following discussion.
The calculated gap at zero temperature $\Delta(0)$ is 6.97 K, which gives
a ratio $2 \Delta/T_c=3.7$, higher than the BCS value.

We now wish to investigate the temperature dependence of the specific heat,
which yields valuable information on the size and nature of the EP coupling.

In a single-band model with a
strong (intermediate) EPI, in the normal state and in the adiabatic
approximation the electronic contribution to the specific heat is determined
from the Eliashberg function $\alpha^{2}(\omega)F(\omega)$ by the expression:
\cite{Grimvall}
\begin{equation}
 C_{N}^{el}(T)    = (2/3)\pi^{2}N(0)k_{B}^{2}T\label{cn}
\end{equation}
\[
\times\left[  1+(6/\pi k_{B}T)\int_{0}^{\infty}f(\omega/2\pi k_{B}%
T)\alpha^{2}(\omega)F(\omega)\omega\right]  ,
\]
where $N(0)$ is a bare DOS per spin at the Fermi energy. The
kernel $f(x)$ is expressed in terms of the derivatives of the digamma function
$\psi(x)$
\begin{equation}
f(x)=-x-2x^{2}\Im\psi^{\prime}( x)-x^{3}\Re\psi
^{\prime\prime}( x). \label{func}%
\end{equation}

At low temperatures the specific heat has the well known asymptotic form:
$C_{N}^{el}(T\rightarrow0)=(1+\lambda)\gamma_{0}T$,
where $\lambda$
is the
electron-phonon coupling constant, and $\gamma_{0}=2\pi^{2}k_{B}^{2}N(0)/3$ is
the specific heat coefficient for noninteracting electrons. At higher
temperatures the specific heat differs from this trivial expression. Below
$T_{c}$ the difference in free energies, $F_{\mathrm{N}}$ and $F_{\mathrm{S}}$,
of the superconducting and normal state
is given by:%

\begin{equation}
-\frac{F_{N}-F_{S}}{\pi N(0)T}=
\label{free}%
\end{equation}

\[
=\left\{ \sum\limits_{n=-\omega_{c}}^{\omega_{c}}%
\begin{array}
[c]{c}%
|\omega_{n}|(Z^{N}(\omega_{n})-1)-\frac{2\omega_{n}^{2}[\left(  Z^{S}%
(\omega_{n})\right)  ^{2}-1]+\varphi_{n}^{2}}{|\omega_{n}|+\sqrt{\omega
_{n}^{2}\left(  Z^{S}(\omega_{n})\right)  ^{2}+\varphi_{n}^{2}}}\\
+\frac{\omega_{n}^{2}Z^{S}(\omega_{n})(Z^{S}(\omega_{n})-1)+\varphi_{n}^{2}%
}{\sqrt{\omega_{n}^{2}\left(  Z^{S}(\omega_{n})\right)  ^{2}+\varphi_{n}^{2}}}%
\end{array}
\right\}  ,
\]
where $Z(\omega_{n})$ is a normalization factor, $\varphi_{n}=\Delta
_{n}/Z(\omega_{n})$ is an order parameter, and $\Delta_{n}$ is the gap function
( see the derivations in \cite{mgb2b} and \cite{d}).

The specific heat at temperature, $T$, is then calculated according to:
\begin{equation}
\Delta C_{\mathrm{el}}(T)=T\partial^{2}(F_{N}-F_{S})/\partial T^{2}.
\label{partial}%
\end{equation}
The specific heat jump $\Delta C_{\mathrm{el}}(T_{c})$ at $T=T_{c}$ is
determined by the coefficient $\beta=T_{c}\Delta C_{\mathrm{el}}(T_{c})/2$
of a second order expansion
$F_{N}-F_{S}=\beta t^{2}$, where $t=(T_{c}-T)/T_{c}$.

\begin{figure}
[ptb]
\begin{center}
\includegraphics[
height=2.5607in,
width=3.3217in
]%
{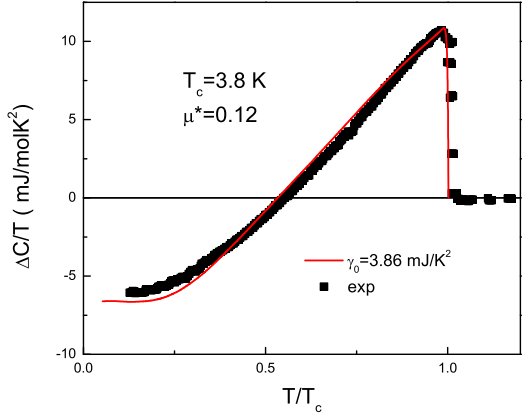}%
\caption{The specific heat of the LaNiASO$_{0.9}$F$_{0.1}$ compound. The
solid(red) line shows results of the calculations. Black dots correspond to
experimental results from Ref. \cite{LNAO:exp:li}}%
\label{spNi}%
\end{center}
\end{figure}

For the comparison with experiment we have considered the data in
Ref.~\cite{LNAO:exp:li}. The anomaly clearly visible at $T_{c}$ in the zero-field data is
suppressed by a magnetic field of 10 Tesla. In figure \ref{spNi}
the difference $\Delta C_{p}=C_{p}(0 \mathrm{Tesla})-C_{p}(10 \mathrm{Tesla})$
is displayed as symbols.%
The specific heat in LaNiAsO$_{0.9}$F$_{0.1}$ was calculated using the isotropic
spectral Eliashberg function $\alpha^{2}(\omega)F(\omega)$ shown in Fig.~\ref{fig:figeph}.
The calculated specific heat at $T_{c}$ is $\gamma^{\mathrm{N}%
}(0)=1.72\gamma_{0}=6.64$ mJ/mol\thinspace K$^{2}$ with $\gamma_{0}$=3.86
mJ/mol\thinspace K$^{2}$ from the band structure calculations. This value is
close to the experimental one of 6.14 reported in Ref.\cite{LNAO:exp:li}
(see Ref.\cite{foot:sp}). The specific heat
jump at $T_{c}$ equals $\Delta C\simeq11.1$ mJ/mol K, which is comparable the
experimental values \cite{LNAO:exp:li}. This gives $\Delta C/(\gamma
^{\mathrm{N}}(0)T_{c})\simeq1.67$ slightly larger than the BCS value of 1.43
which corresponds to the intermediate coupling. To estimate the specific heat
jump we can also apply the semiempirical expression by Carbotte \cite{Carb}%
\[
\Delta C/T_{c}=1.43\gamma_{0}(1+\lambda)\left[  1+(\frac{T_{c}}{\omega_{\ln}%
})^{2}\ln(\frac{\omega_{\ln}}{3T_{c}})\right]  .
\]
With $\omega_{\ln}=96$ K and with $T_{c}=3.8$ K, we have
$\Delta C/T_{c}=11.2$ mJ/molK$^{2}$, which compares well with the full numerical solution.
The difference $\Delta C_{\mathrm{el}}(T)=C_{\mathrm{el}}^{S}%
(T)-C_{\mathrm{el}}^{N}(T)$ , shown in Figure~\ref{spNi} as solid line, shows
a very good agreement with the experimental data. We would like to emphasize here that no
fitting is involved in the theoretical calculations. The only free parameter
which is in the Coulomb matrix element $\mu^*$, which was  determined by the
experimental $T_{c}$.

Here we have to point out that the specific
heat jumps in multiband systems ( or other anisotropic ones) is sufficiently
smaller than in one-band superconductors ( see, discussions in \cite{mgb2b} and
\cite{d}, and some criteria in \cite{SD}).
These results show that $T_{c}$ and the specific heat in LaNiASO$_{0.9}$F$_{0.1}$,
in contrast to LaFeAsO$_{0.9}$F$_{0.1}$, can be described in the framework of the
standard single-band approach without the need of exotic mechanisms.
To further support this conclusion, we observe that recent de-Haas van Alphen
experiemnts on an other Ni pnictide, BaNi$_2$P$_2$,
observe a band structure which is in close agreement with DFT
calculations, with an average effective mass renormalization of $\sim 1.8$,
which implies an EP coupling constant $\lambda=0.8$, in good agreement with our
calculations~\cite{BN2P2:exp:terashima}.

\section{Discussion:}
To summarize the results of the previous sections, we have found that
linear response calculations of the electron-phonon coupling yield
rather different results for the two superconducting members of
the LaO$M$As family of pnictide superconductors.

For the Fe compound the
value of the total EP coupling constant $\lambda=0.21$ is
much lower than in normal EP superconductors (for example, $\lambda=0.44$
in Al, where $T_c$ is 1.3 K), and even the inclusion of multiband
effects cannot explain the $T_c$=26 K observed in doped samples.

For the Ni compound the coupling constant is much higher
($\lambda=0.72$); its normal and superconducting state properties can be well described
by standard, single-band Migdal-Eliashberg theory.
The values of the gap ratio
$2 \Delta/T_c=3.7$ and specific heat jump $\Delta C/\gamma^N(0)T_c=1.67$
are larger than what predicted by BCS theory (3.52 and 1.43) respectively.

The picture that emerges from our calculations is that of a family
of rather standard EP superconductors (Ni-based), opposed to a family
of ``exotic'' superconductors (Fe-based), which is supported by several
experimental evidences.

The most important issue is the magnetic ground state
of the superconducting samples.
The Ni parent compounds are standard metals, which superconduct at low
T $<$ 5 K~\cite{LNAO:tc:watanabe,LNAO:exp:li,LNAO:exp:fang,BN2A2:exp:ronning,LNPO:exp:watanabe,BN2P2:exp:terashima} 
\footnote{A possible SDW transition has been observed at $\sim 66$ K
in BaNi$_2$As$_2$ in Ref. \cite{BN2A2:exp:ronning}};
doping (holes or electrons) does not change $T_c$ or normal state properties
dramatically. The calculated coupling constants $\lambda=0.58-0.76$~\cite{LNPO:DFT:subedi,BN2A2:DFT:subedi} can
well explain the experimental $T_c$ and, as we have shown in the
previous section, also thermodynamic properties and de-Haas-van-Alphen data.

On the other hand, in the Fe compounds superconductivity only
appears by doping a parent compound which is an AFM metal.
At the time when our calculations presented in Ref.~\cite{LFAO:DFT:boeri}
were performed, the only available experimental data showed
that doping suppressed the static AFM order in the superconducting samples.~\cite{LFAO:tc:kamihara}
Therefore, we assumed that, as in the Ni superconductors, also in Fe superconductors the normal state is non-magnetic.

In the last few months, experiments have shown that the ground state of
the superconducting samples may also be magnetic, but with fluctuating
(dynamic) moments.
For a more complete discussion of this subject,
see the review by Mazin and Schmalian in this issue.
Such an arrangement is not describable by DFT theory, therefore it is
not possible to estimate what its effect on the EP coupling would be.

It is also hard to compare our results with experimental data. There are
no direct measurements of the EP coupling in literature, although
some experiments (ARPES, penetration depth, specifit heat) indicate
some retarded electron-boson interaction, with a coupling constant
$\lambda^B \approx 0.5-1.5$; however, the total coupling could be due to
other bosonic excitations.

On the other hand, a few measurements of phonon spectra are available in literature~\cite{phonon,christianson,fukuda,mittal1,mittal2,reznik}.
In general, there is a good agreement between experimental phonon
frequencies and non-spin polarized calculations,
except for the intermediate frequency
Fe-As (and As-As) modes, which are lower in experiment than in calculations.
See for example Ref.~\cite{christianson},
where Inelastic Neutron Scattering data are compared to our PDOS.
An empirical way to reconcile experiment and theory, by reducing
the Fe-As force constant, was proposed in Ref.~\cite{fukuda}.
It was later shown that
the inclusion of a static AFM order leads indeed to a softening the Fe-As spring constant,
and improves the agreement of the predicted crystal structure and phonon frequencies with
experiment~\cite{Zbiri};
however, this cannot explain the softening of  $c$-polarized
As modes, which form a distinct peak at 20 meV, at an energy 20 $\%$
lower than predicted by calculations, which has been observed both
in 1111 and 122 samples, and has been attributed to anomalous $e-ph$
coupling~\cite{mittal1,reznik}.

It was realized very early that Fe pnictide show a strong magneto-elastic coupling between Fe moments and As out-of-plane modes~\cite{LFAO:DFT:singh,LFAO:DFT:boeri,LFAO:DFT:yin}; in Ref.~\cite{soler}
it was proposed that this leads to an increased EP coupling.
In principle, also many body effects could increase the coupling constant
beyond the LDA value.

In conclusion, on the basis of our results we can exclude that standard
EP coupling theory alone can cause the observed $T_c$ in Fe pnictides;
however, this does not mean that the phonons play no role in the superconducting pairing, as
they might enhance or reduce the pairing due to other mechanisms.
If this is the case, it is not surprising to observe a finite Fe
isotope effect on $T_c$.

On the othere hand, LaONiAs represents a nice example of a single-gap, strong-coupling EP
superconductor. It is important to stress that the difference between the
two compounds can be traced back essentially to a different filling of the
same complicated, non-magnetic band structure, which derives from a non trivial
hybridization between $M$ and pnictogen atoms.

\section*{Acknowledgements:}
We wish to acknwoledge useful discussion at various stages of the
present work with Ole K. Andersen, Igor I. Mazin, Aleksander N. Yaresko,
Reinhard K. Kremer, Alessandro Toschi and Giorgio Sangiovanni.
\appendix

\section{Computational Details:}
\label{details}
For the atom-projected band and DOS plots in Fig.~\ref{fig:fatFe}-\ref{fig:fatNi} we
employed the full-potential
LAPW method~\cite{DFT:LAPW:andersen} as implemented in the
Wien2k code~\cite{DFT:WIEN2k}.
Calculations of phonon spectra, EP  coupling
and structural relaxations were performed using
planewaves and pseudopotentials with {\sc QUANTUM-espresso}~\cite{DFT:PWSCF}.
We employed ultrasoft Vanderbilt pseudopotentials~\cite{Vanderbilt},
 with a cut-off of 40 Ryd for the wave-functions, and 320 Ryd for
the charge densities.
The $\mathbf{k}$-space integration for the electrons
 was approximated by a summation
 over a 8 8 4 uniform grid in reciprocal space, with a Gaussian smearing
of 0.02 Ryd for self-consistent cycles and relaxations; a much finer
(16 16 8) grid was used for evaluating DOS and
EP linewidths.
Dynamical matrices and EP linewidths were calculated
on a uniform 442 grid in $\mathbf{q}$-space; phonon dispersions and DOS
were then obtained by Fourier interpolation of the dynamical matrices,
and the Eliashberg function by summing over individual linewidths and phonons.
To check the effect of nesting on the EP linewidhts, we also calculated
selected $\mathbf{q}$ points on a 882 grid.

Whenever possible, we cross-checked the results given by the two codes
and found them to be in close agreement; for consistency, we used
the same GGA-PBE exchange-correlation potential in both cases~\cite{DFT:PBE}.



\begin{thebibliography}{100}
\bibitem{foot:sp}The value $\gamma^N(0)=6.14$ mJ/mol K$^2$ is obtained from the experimental
$\gamma^N=7.3$ mJ/molK$^2$ in Ref.~\cite{LNAO:exp:li}, considering that the
superconducting volume fraction is 84 $\%$.
\end{thebibliography}
\end{document}